\documentclass[12pt]{article} 
\usepackage{epsfig}
\usepackage{a4}
\usepackage{latexsym}
\usepackage{cite}
\usepackage{color}

\usepackage{pslatex}
\usepackage[latin1]{inputenc}
\usepackage[T1]{fontenc}

\newcommand{\hspn}{{\hspace{-5mm}}}

\newcommand{\beq}{\begin{equation}}
\newcommand{\eeq}{\end{equation}}
\newcommand{\bea}{\begin{eqnarray}}
\newcommand{\eea}{\end{eqnarray}}
\newcommand{\nn}{\nonumber}
\newcommand{\MSb}{$\overline{\mbox{MS}}$}
\newcommand{\as}{\alpha_{\rm s}}
\newcommand{\ar}{a_{\rm s}}

\newcommand{\wK}{{\widetilde K}}
\newcommand{\wc}{{\tilde c}}
\newcommand{\wg}{{\widetilde \gamma}}

\def\FR#1#2{\mbox{\small{$\displaystyle\frac{#1}{#2}$}}}

\def\S(#1){{{S}_{#1}}}
\def\Ss(#1,#2){{{S}_{#1,#2}}}
\def\Sss(#1,#2,#3){{{S}_{#1,#2,#3}}}
\def\Ssss(#1,#2,#3,#4){{{S}_{#1,#2,#3,#4}}}
\def\Sssss(#1,#2,#3,#4,#5){{{S}_{#1,#2,#3,#4,#5}}}

\begin{document}

\setlength{\parskip}{0.2cm}
\setlength{\baselineskip}{0.53cm}

\def\Qs{{Q^{\, 2}}}
\def\ca{{C_A}}
\def\cas{{C^{\: 2}_A}}
\def\cath{{C^{\: 3}_A}}
\def\cafo{{C^{\: 4}_A}}
\def\cafi{{C^{\: 5}_A}}
\def\cf{{C_F}}
\def\cfs{{C^{\: 2}_F}}
\def\cft{{C^{\: 3}_F}}
\def\cff{{C^{\: 4}_F}}
\def\nf{{n^{}_{\! f}}}
\def\nfs{{n^{\,2}_{\! f}}}
\def\nft{{n^{\,3}_{\! f}}}
\def\nff{{n^{\,4}_{\! f}}}
\def\camcf{(\ca-\cf)}
\def\camtcf{(\ca-2\*\,\cf)}
\def\dfourAAna{\FR{d_A^{\,abcd}d_A^{abcd}}{n_A^{}}}
\def\dfourRRna{\FR{d_F^{\,abcd}d_F^{abcd}}{n_A^{}}}
\def\dfourRAna{\FR{d_F^{\,abcd}d_A^{abcd}}{n_A^{}}}
\def\dfourRRnr{\FR{d_F^{\,abcd}d_F^{abcd}}{n_R^{}}}
\def\dfourRAnr{\FR{d_F^{\,abcd}d_A^{abcd}}{n_R^{}}}
\newcommand{\colourcolour}[1]{{\color{blue}{#1}}}

\def\pip#1{{\pi^{\,#1}}}
\def\z#1{{\zeta_{\,#1}}}
\def\zss{\zeta_{\,2}^{\,2}}
\def\zst{\zeta_{\,2}^{\,3}}

\def\arp#1{a_{\rm s}^{\,#1}}
\def\bb#1{\beta_{#1}}

\def\muRs{\mu_{\mbox{\footnotesize \sc r}}^{\,2}}

\begin{titlepage}

\noindent
TTP$\,$17-045 \hspace*{\fill} November 2017\\
LTH 1145 \\
\vspace{2.0cm}
\begin{center}
\Large
{\bf Absence of $\pi^2$ terms in physical anomalous dimensions in DIS:
\\[1mm]
verification and resulting predictions}\\
\vspace{2cm}
\Large
J. Davies$^{\:\!a}$ and A. Vogt$^{\:\!b}$\\
\vspace{1.5cm}
\normalsize

{\it $^a$Institute for Theoretical Particle Physics, 
Karlsruhe Institute of Technology\\ 
\vspace{1mm} D-76128 Karlsruhe, Germany}\\ 
\vspace{0.5cm}
{\it $^b$Department of Mathematical Sciences, University of Liverpool\\
\vspace{1mm} 
Liverpool L69 3BX, United Kingdom}\\[2.5cm]
\vfill
\large {\bf Abstract}
\vspace{-0.2cm}
\end{center}
We study the higher-order corrections to structure functions in inclusive 
deep-inelastic scattering (DIS) in massless perturbative QCD, in the context 
of the conjectured absence of even-$n$ values of the Riemann zeta-function 
$\zeta_{\,n}$, i.e., of powers of $\pip2$, in Euclidean physical quantities.  
We provide substantial additional support for this conjecture by demonstrating
that it holds, as far as it can be tested by the results of diagram 
computations, for the physical anomalous dimensions of structure functions
at the fourth and fifth order in the strong coupling constant $\as$. 
The conjecture is then employed to predict hitherto unknown $\z4$ and $\z6$ 
contributions to the anomalous dimensions for parton distributions and to the 
coefficient functions for the longitudinal structure function $F_{\:\!L}$.

\vspace{1.0cm}
\end{titlepage}
%

\noindent
As far as they are presently known, the anomalous dimensions 
-- i.e., the even-$N$ or odd-$N$ Mellin moments of the splitting functions --
for the scale dependence (evolution) of parton distributions can be expressed
in terms of rational numbers and integer-$n$ values $\zeta_{\,n}^{}$ of the Riemann 
$\zeta$-function. 
The~same holds for the $N$-space coefficient functions for inclusive 
lepton-hadron deep-inelastic scattering (DIS) via the exchange of a boson with 
spacelike four-momentum $q$, i.e., $q^{\:\!2} \equiv -\Qs < 0$.
Three-loop calculations at $N \leq 14$ of these quantities were performed 
in refs.~\cite{3loopN1,3loopN2,3loopN3}. The corresponding all-$N$ expressions 
were derived in refs.~\cite{P3loop1} for the anomalous dimensions 
(see also ref.~\cite{P3loop2} for the helicity-dependent case) 
and in refs.~\cite{C3loop1} for the most important structure functions in DIS.

It is an old observation that the above `spacelike' quantities do not include 
terms linear in $\z2 = \pip2/6$ 
\footnote{This does not hold for the moments of the splitting functions for 
 fragmentation distributions \cite{P2frag} and the coefficient functions for 
 semi-inclusive $e^+e^-$ annihilation via a boson with a timelike 
 four-momentum. It also does not hold for `unnatural' (non-OPE) moments of 
 DIS quantities such as the odd moments of the photon-exchange $F_{\,2}$
 \cite{MRV07}.}.
Terms with $\zss$ or $\,\z4 = 2/5\, \zss = \pip4/90$ do occur in the three-loop
coefficient functions in DIS (with the exception of the longitudinal structure
function $F_{\:\!L\,}$) \cite{3loopN1,3loopN2,3loopN3} and the four-loop 
anomalous dimensions \cite{avLL2016,MRUVV17,RUVVprp}, which together define 
the next-to-next-to-next-to-leading order (N$^3$LO) approximation in 
renormalization-group improved perturbation theory.
The~corresponding N$^4$LO quantities, i.e., the four-loop coefficient 
functions and five-loop anomalous dimensions, include contributions with $\z4$ 
and $\,\z6 = 8/35\, \zst = \pip6/945$ \cite{RUVVprp,RadcTalk,HMRUVVprp}.

Already about 20 years ago, the absence of also $\z4$ in the perturbative 
expansion of spacelike (Euclidean) physical quantities was referred to as an 
empirical rule 
\cite{3loopN2}.  In the standard \MSb\ renormalization scheme, however, this 
rule is violated at N$^4$LO by the scalar quark and gluon correlators that 
enter the hadronic decays of the Higgs boson \cite{HbbggN4LO}.
Very recently, it has been demonstrated in ref.~\cite{JM-no-pi2} that the $\z4$ 
terms in the above quantities vanish after transforming the coupling constant
to the $C$-scheme introduced in ref.~\cite{Cscheme}. This highly non-trivial
cancellation can occur since the five-loop beta function of QCD and its 
gauge-group generalizations include $\z4$ with most colour factors \cite{beta4}.
It has thus been conjectured that all even-$n$ $\zeta$-values, i.e., all
powers of $\pip2$, are absent from all spacelike physical quantities in 
massless perturbative QCD in this scheme~\cite{JM-no-pi2}.

The factorization-scheme dependent anomalous dimensions and coefficient
functions can be combined to form physical anomalous dimensions for structure
functions in DIS. 
In this manner, the above `no-$\pip2$ conjecture' can be {\bf (a)} supported
(or falsified) by quantities not considered in this context so far, and 
{\bf (b)} used to predict new results for higher-order coefficients in the
perturbation series. The latter possibility was, in fact, already mentioned 
(but not followed up) in ref.~\cite{3loopN2}.
In~the present letter, we perform both steps at N$^3$LO and N$^4$LO for the 
non-singlet structure functions $F_{\,2,\rm ns}$ and $F_{\,3}$ and for the 
flavour-singlet system $(F_{\,2},\,F_{\,\phi})$. 
Here $F_\phi$ is the structure function for DIS via the exchange of a scalar 
that (like the Higgs boson in the limit of a heavy top quark and $\nf$ 
effectively massless flavours \cite{HGGeff}) couples directly only to gluons
\cite{FP82}. 
We finally also address physical anomalous dimensions for $F_{\:\!L}$.

The physical anomalous dimensions $K$ can be obtained by considering the 
scaling violations $dF/d\ln \Qs$ of the (vector of) $N$-space structure 
functions $F$, using the evolution equations for the parton distributions $q$ 
and then expressing these in terms of the structure functions, viz
\beq
\label{Kdef}
  \frac{d F}{d \ln \Qs}  \:=\:
  \frac{d}{d \ln \Qs} \, ( C \, q) \:=\:
  \frac{d\, C}{d \ln \Qs}\, q \, - \, C\, \gamma \, q \:=\:
  \Big( \beta\: \frac{d\, C}{d \ar}\, - \, C\, \gamma \Big)\, C^{\,-1} F 
  \:\equiv\: K\, F
\; .
\eeq
Here $C$ represents the coefficient functions, $\gamma$ the anomalous 
dimensions (we use the standard convention $\gamma = - P$ for their relation
to the moments of the splitting functions), $\ar \equiv \as/4 \pi$ the 
reduced strong coupling constant and $\beta$ the beta function of QCD and 
its generalizations. We~have suppressed here, as in many instances below, the 
dependence of $C$, $\gamma$ and $K$ on $N$ for brevity. Our~notation
(and normalization) for the expansion coefficients of these quantities is
\bea
  C \:=\: 1 + \sum_{\ell=1} \arp{\ell} \, c^{\,(\ell)} 
\; , \;\;\;
  \{ \gamma\,,\,K \} \:=\: \sum_{\ell=1} \arp{\ell} \,
  \{ \gamma\,,\,K \}^{\,(\ell-1)}
\; .
\eea

For the non-singlet cases, eq.~(\ref{Kdef}) is simplified by 
$C\, \gamma\: C^{\,-1} \!= \gamma$, thus only the N$^n$LO anomalous 
dimensions $\gamma^{\,(n)} $
(together with the coefficient functions at this and all previous orders)
enter the physical anomalous dimensions at N$^n$LO. 
In this case one has $K = -\gamma\,$ for those colour factors, such as $\cff$
at order $\arp4$, that cannot be generated by multiplying the beta function 
and powers of the coefficient functions.
For the system $(F_{\,2},\,F_{\,\phi})$, the quantities $F$, $C$, $\gamma$ and 
$K$ represent the matrices 
\bea
\label{F2Fphi}
  F \:=\:
  \Big( \begin{array}{c}  F_{\,2} \\[-1mm] F_{\,\phi} \end{array} \Big)
\; , \;\;\;
  C \:=\:
  \Big(  \begin{array}{cc} C_{2,\rm q}^{} & C_{2,\rm g}^{} \\[-0.5mm]
  C_{\phi\!,\,\rm q} & C_{\phi\!,\,\rm g} \end{array} \Big)
\; , \;\;\;
 K \:=\:
 \Big( \begin{array}{cc} K_{22} & K_{2\phi} \\[-0.5mm]
  K_{\phi 2} & K_{\phi\phi} \end{array} \Big)
\; , \;\;\;
 \gamma \:=\:
 \Big( \begin{array}{cc} \gamma_{\,\rm qq} & \gamma_{\,\rm qg} \\[-0.5mm]
 \gamma_{\,\rm gq} & \gamma_{\,\rm gg} \end{array} \Big)
\; .
\eea
The non-singlet expansion coefficients of $K$ have been written down to order 
$\arp5$, the maximal order of the present study, in eqs.~(2.7) and (2.8) of 
ref.~\cite{NVphys} for the choice $\muRs = \Qs$ of the renormalization 
scale already employed in eq.~(\ref{Kdef}). 
The generalization to $\,\muRs \neq \Qs$, see eq.~(2.9) of ref.~\cite{NVphys}, 
does not provide new information. 
Corresponding matrix-notation results for systems like $(F_{\,2},\,F_{\,\phi})$
have been given to $\arp4$ in eq.~(2.25) of ref.~\cite{C3loop2}; their 
extension to the $\arp5$ terms is straightforward.

Only a small part of the complete expressions is required for the $\z4$ and
$\z6$ contributions. Consequently, the $\z4$ parts of the $\arp4$ (N$^3$LO)
physical anomalous dimensions considered for now take the simple form
\beq
\label{Kns3z4}
  \wK_{2,\rm ns}^{\:(3)} \:=\: \mbox{}
  - \wg_{\,\rm ns}^{\:+(3)} - 3\:\! \bb0\, \wc_{2,\rm ns}^{\:(3)} 
\; , \qquad
    \wK_{3}^{\:(3)} \:=\: \mbox{}
  - \wg_{\,\rm ns}^{\:-(3)} - 3\:\! \bb0\, \wc_{3}^{\:(3)}
\eeq
and
\bea
\label{K2p3z4}
  \wK_{22}^{\:(3)} &\!\!=\!\!\!& \mbox{}
  - \wg_{\,\rm qq}^{\:(3)} - 3\:\! \bb0\, \wc_{2,\rm q}^{\:(3)}
  \,-\, \gamma_{\,\rm gq}^{\,(0)} \, \wc_{2,\,\rm g}^{\:(3)}
  \,+\, \gamma_{\,\rm qg}^{\,(0)} \, \wc_{\phi,\,\rm q}^{\:(3)}
\; , \nn \\[1mm]
  \wK_{2\phi}^{\:(3)} &\!\!=\!\!\!& \mbox{}
  - \wg_{\,\rm qg}^{\:(3)} - 3\:\! \bb0\, \wc_{2,\rm g}^{\:(3)}
  \,-\, \gamma_{\,\rm qg}^{\,(0)}
     \Big( \wc_{2,\rm q}^{\:(3)} - \wc_{\phi,\,\rm g}^{\:(3)} \Big)
  \,-\, \wc_{2,\rm g}^{\:(3)} 
     \Big( \wg_{\,\rm qq}^{\:(3)} - \wg_{\,\rm gg}^{\:(3)} \Big)
\; , \nn \\[1mm]
  \wK_{\phi 2}^{\:(3)} &\!\!=\!\!\!& \mbox{}
  - \wg_{\,\rm gq}^{\:(3)} - 3\:\! \bb0\, \wc_{\phi,\rm q}^{\:(3)}
  \,-\, \gamma_{\,\rm gq}^{\,(0)}
     \Big( \wc_{2,\rm q}^{\:(3)} - \wc_{\phi,\,\rm g}^{\:(3)} \Big)
  \,-\, \wc_{\phi,\rm q}^{\:(3)} 
     \Big( \wg_{\,\rm qq}^{\:(3)} - \wg_{\,\rm gg}^{\:(3)} \Big)
\; , \nn \\[1mm]
  \wK_{\phi\phi}^{\:(3)} &\!\!=\!\!\!& \mbox{}
  - \wg_{\,\rm gg}^{\:(3)} - 3\:\! \bb0\, \wc_{\phi,\rm g}^{\:(3)}
  \,-\, \gamma_{\,\rm qg}^{\,(0)} \, \wc_{\phi,\,\rm q}^{\:(3)}
  \,+\, \gamma_{\,\rm gq}^{\,(0)} \, \wc_{2,\,\rm g}^{\:(3)}
\eea
in terms of the four-loop anomalous dimensions and three-loop coefficient 
functions. The latter~are completely known \cite{C3loop1,C3loop2}.
Here and below a tilde above a quantity indicates the coefficient of $\z4$.
If the no-$\pip2$ conjecture is correct, then the left-hand sides of 
eqs.~(\ref{Kns3z4}) and eqs.~(\ref{K2p3z4}) vanish for all~$N$.

Diagram calculations of the N$^3$LO non-singlet anomalous dimensions up to 
$N=16$ (and all-$N$ expressions in the limit of a large number of colours 
$n_{\rm c}$) have been presented in ref.~\cite{MRUVV17}. 
These results are sufficient for determining the all-$N$ expressions for 
$\,\wg_{\,\rm ns}^{\;\pm(3)}(N)$, which are found to read
\beq
\label{gnspm3z4}
\wg_{\,\rm ns}^{\;\pm(3)} =\,
8\*\,\colourcolour{\cf\*(\ca-\cf)\*\,\beta_0}\* \Big[
  9\*\,\colourcolour{(\ca-2\*\,\cf)} \* \Big( \,
       \FR{5}{4}
     -  ( \eta + \eta^2 ) \* (\pm1)^N
     +  2 \*\, \S(-2)
    \Big)
  - 3\,\*\colourcolour{\nf} \* \Big( \,
       \FR{3}{2}
     +  \eta
     -  2 \* \S(1) 
    \Big)\Big]
\eeq
with $\eta \equiv D_0 - D_1 \equiv 1/N - 1/(N\!+\!1)$. Here and below the
argument $N$ of all harmonic sums $S_{\vec{w}}(N)$ \cite{Hsums} is suppressed for brevity.
For $\wg_{\,\rm ns}^{\:+(3)\!}$, the anomalous dimension for flavour 
differences of quark-antiquark sums, eq.~(\ref{gnspm3z4}) is valid at even $N$. 
For its quark-antiquark difference counterpart $\wg_{\,\rm ns}^{\:-(3)}$ it 
provides the odd-$N$ values with, of course, 
$\wg_{\,\rm ns}^{\:-(3)}(N\!=\!1) \,=\, 0$ as required by fermion number
conservation. 
Using eq.~(\ref{gnspm3z4}) in eq.~(\ref{Kns3z4}) we indeed~find
\beq
\label{Knsz4=0}
  \wK_{2,\rm ns}^{\:(3)} \:=\: \wK_{3}^{\:(3)} \:=\: 0 
  \quad \mbox{for all even$/$odd $\,N$} \;.
\eeq

The diagram calculations of four-loop singlet splitting functions, performed
along the lines of refs.~\cite{3loopN1,3loopN2,3loopN3} with the {\sc Forcer} 
program \cite{Forcer} for massless four-loop self-energy integrals,
have been completed so far only at $N=2$ and $N=4$. The result of the hardest 
of these calculations, $\gamma_{\,\rm gg}^{\:(3)}(N\!=\!4)$, has been given in 
eq.~(2.1) of ref.~\cite{avLL2016}; the other results will be presented in 
ref.~\cite{RUVVprp}. Inserting these results into eq.~(\ref{K2p3z4}), we find
\beq
\label{Kijz4=0}
  \wK_{22}^{\:(3)} \:=\: \wK_{2\phi}^{\:(3)} \:=\: 
  \wK_{\phi 2}^{\:(3)} \:=\: \wK_{\phi\phi}^{\:(3)} \:=\: 0
\eeq
for $N=2$ and $N=4$, thus verifying the no-$\pip2$ conjecture also in the 
singlet sector. Imposing eq.~(\ref{Kijz4=0}) for all even $N$, we can now 
predict the complete results for all four quantities 
$\,\wg_{\,\rm ik}^{\:(3)}(N)$ in eq.~(\ref{F2Fphi}): 
\bea
  \label{gps3z4}
  \wg_{\,\rm ps}^{\:(3)} &\!\!=\!\!&\vphantom{\FR{1}{1}}
16\*\,\colourcolour{\cf\*\,\camcf} \*\, \Big[
       \colourcolour{\nfs}\*\, \Big(
          15\*\,\eta
          + 10\*\,\eta^2
          - 20\*\,\nu
          \Big)
       + \colourcolour{\cf\*\,\nf} \*\, \Big(
          138\*\,\nu
          - 72\*\,\nu^2
          - 117\*\,\eta
\nn\\[-1mm]&&\hspn\vphantom{\FR{1}{1}}
          - 87\*\,\eta^2
          - 18\*\,\eta^3
          \Big)
       +\, \colourcolour{\camcf\*\,\nf}\*\, \Big(
          114\*\,\nu
          - 72\*\,\nu^2
          - \FR{195}{2}\*\,\eta
          - 69\*\,\eta^2
          - 12\*\,\eta^3
          + [\, 24\*\,\nu
\nn\\[-1mm]&&\hspn\vphantom{\FR{1}{1}}
          - 18\*\,\eta 
          - 12\*\,\eta^2 \,] \*\,\S(1)
          \Big)
\Big]
\:\:, \\[3mm]
  \label{gqg3z4}
  \wg_{\,\rm qg}^{\:(3)} &\!\!=\!\!&\vphantom{\FR{1}{1}}
16\*\,\colourcolour{\camcf} \*\, \Big[
       \colourcolour{\camcf \*\, \nfs}  \*\,  \Big(
          22 \*\, D_0
          - \FR{16}{3} \*\, D_{-1}
          - 4 \*\, D_0^2
          + 6 \*\, D_0^3
          - 50 \*\, D_1
          + 10 \*\, D_1^2
\nn\\[-1mm]&&\hspn\vphantom{\FR{1}{1}}
          - 12 \*\, D_1^3
          + \FR{109}{3} \*\, D_2
          + 16 \*\, D_2^2
          + [\, 4 \*\, D_0 - 8 \*\, D_1 + 8 \*\, D_2 \,] \*\, \S(1)
          \Big)
       + \colourcolour{\camcf^2 \*\, \nf}  \*\,  \Big(
          \FR{122}{3} \*\, D_{-1}
\nn\\[-1mm]&&\hspn\vphantom{\FR{1}{1}}
          - 8 \*\, D_{-1}^2
          - \FR{91}{2} \*\, D_0
          - 21 \*\, D_0^2
          - 18 \*\, D_0^3
          + 208 \*\, D_1
          - 96 \*\, D_1^2
          + 36 \*\, D_1^3
          - \FR{659}{3} \*\, D_2
          - 114 \*\, D_2^2
\nn\\[-1mm]&&\hspn\vphantom{\FR{1}{1}}
          - 24 \*\, D_2^3
          + [\, 12 \*\, D_{-1} + 13 \*\, D_0 - 24 \*\, D_0^2 + 46 \*\, D_1 
            - 48 \*\, D_1^2 - 91 \*\, D_2 - 36 \*\, D_2^2 \,] \*\, \S(1)
          + [\, 24 \*\, D_1 
\nn\\[-1mm]&&\hspn\vphantom{\FR{1}{1}}
            - 12 \*\, D_0 - 24 \*\, D_2 \,]\*\, \Ss(1,1)
          + [\, 6 \*\, D_0 - 12 \*\, D_1 + 12 \*\, D_2 \,] \*\, \S(2)
          \Big)
       + \colourcolour{\cf \*\, \nfs}  \*\,  \Big(
          \FR{49}{2} \*\, D_0
          - 8 \*\, D_{-1}
          + D_0^2
\nn\\[-1mm]&&\hspn\vphantom{\FR{1}{1}}
          + 6 \*\, D_0^3
          - 70 \*\, D_1
          + 20 \*\, D_1^2
          - 12 \*\, D_1^3
          + 55 \*\, D_2
          + 24 \*\, D_2^2
          + [\, 2 \*\, D_0 - 4 \*\, D_1 + 4 \*\, D_2 \,]\*\, \S(1)
          \Big)
\nn\\[-1mm]&&\hspn\vphantom{\FR{1}{1}}
       + \colourcolour{\cf \*\, \camcf \*\, \nf}  \*\,  \Big(
          \FR{268}{3} \*\, D_{-1}
          - 16 \*\, D_{-1}^2
          - 121 \*\, D_0
          - 55 \*\, D_0^2
          - 36 \*\, D_0^3
          + 515 \*\, D_1
          - 218 \*\, D_1^2
\nn\\[-1mm]&&\hspn\vphantom{\FR{1}{1}}
          + 72 \*\, D_1^3
          - \FR{1531}{3} \*\, D_2
          - 252 \*\, D_2^2
          - 48 \*\, D_2^3
          + [\, 16 \*\, D_{-1} - \FR{41}{2} \*\, D_0 - 27 \*\, D_0^2 
            + 122 \*\, D_1 - 54 \*\, D_1^2
\nn\\[-1mm]&&\hspn\vphantom{\FR{1}{1}}
          - 144 \*\, D_2 - 48 \*\, D_2^2 \,]\*\, \S(1)
          + [\, 24 \*\, D_1 - 12 \*\, D_0 - 24 \*\, D_2 \,]\*\, \Ss(1,1)
          + [\, 12 \*\, D_1 - 6 \*\, D_0 - 12 \*\, D_2 \,]\*\, \S(-2)
\nn\\[-1mm]&&\hspn\vphantom{\FR{1}{1}}
          + [\,6 \*\, D_0 - 12 \*\, D_1 + 12 \*\, D_2 \,]\*\, \S(2)
          \Big)
       + \colourcolour{\cfs \*\, \nf}  \*\,  \Big(
          \FR{146}{3} \*\, D_{-1}
          - 8 \*\, D_{-1}^2
          - \FR{299}{4} \*\, D_0
          - 34 \*\, D_0^2
          - 21 \*\, D_0^3
\nn\\[-1mm]&&\hspn\vphantom{\FR{1}{1}}
          + \FR{629}{2} \*\, D_1
          - 131 \*\, D_1^2
          + 42 \*\, D_1^3
          - \FR{854}{3} \*\, D_2
          - 138 \*\, D_2^2
          - 24 \*\, D_2^3
          + [\, 4 \*\, D_{-1} - \FR{13}{2} \*\, D_0 - 9 \*\, D_0^2
\nn\\[-1mm]&&\hspn\vphantom{\FR{1}{1}}
             + 40 \*\, D_1 - 18 \*\, D_1^2 - 53 \*\, D_2 
             - 12 \*\, D_2^2 \,] \*\, \S(1)
          + [\, 6 \*\, D_0 - 12 \*\, D_1 + 12 \*\, D_2 \,]\*\, \S(-2)
          \Big)
\Big]
\:\:,\\[3mm]
  \label{ggq3z4}
  \wg_{\,\rm gq}^{\:(3)} &\!\!=\!\!&\vphantom{\FR{1}{1}}
16\*\,\colourcolour{\camcf} \*\, \Big[
       \colourcolour{\cf \*\, \nfs}  \*\,  \Big(
          \FR{16}{3} \*\, D_{-1}
          - \FR{16}{3} \*\, D_0
          + \FR{8}{3} \*\, D_1
          \Big)
       + \colourcolour{\cf \*\, \camcf \*\, \nf}  \*\,  \Big(
          \FR{226}{3} \*\, D_0
\nn\\[-1mm]&&\hspn\vphantom{\FR{1}{1}}
          - \FR{85}{3} \*\, D_{-1}
          - 18 \*\, D_0^2
          + 12 \*\, D_0^3
          - \FR{167}{3} \*\, D_1
          - 18 \*\, D_1^2
          - 6 \*\, D_1^3
          + [\, 4 \*\, D_0 - 4 \*\, D_{-1} - 2 \*\, D_1 \,]\*\, \S(1)
          \Big)
\nn\\[-1mm]&&\hspn\vphantom{\FR{1}{1}}
       + \colourcolour{\cf \*\, \camcf^2}  \*\,  \Big(
          88 \*\, D_1
          + 34 \*\, D_1^2
          + 6 \*\, D_1^3
          - 25 \*\, D_{-1}
          - 12 \*\, D_{-1}^2
          - 92 \*\, D_0
          + 50 \*\, D_0^2
\nn\\[-1mm]&&\hspn\vphantom{\FR{1}{1}}
          - 12 \*\, D_0^3
          - 4 \*\, D_2
          + [\, 24 \*\, D_0 - 24 \*\, D_{-1} - 12 \*\, D_1 \,] \*\, \Ss(1,1)
          + [\, 12 \*\, D_{-1} - 12 \*\, D_0 + 6 \*\, D_1 \,] \*\, \S(2)
\nn\\[-1mm]&&\hspn\vphantom{\FR{1}{1}}
          + [\, 37 \*\, D_{-1}
            + 12 \*\, D_{-1}^2 - 20 \*\, D_0 + 10 \*\, D_1 
            + 4 \*\, D_2 \,]\*\, \S(1)
          \Big)
       + \colourcolour{\cfs \*\, \nf}  \*\,  \Big(
          8 \*\, D_{-1}^2
          - 55 \*\, D_{-1}
          + \FR{190}{3} \*\, D_0
\nn\\[-1mm]&&\hspn\vphantom{\FR{1}{1}}
          + 8 \*\, D_0^2
          + 12 \*\, D_0^3
          - \FR{181}{6} \*\, D_1
          - 5 \*\, D_1^2
          - 6 \*\, D_1^3
          + \FR{8}{3} \*\, D_2
          + [\, 8 \*\, D_{-1} - 8 \*\, D_0 + 4 \*\, D_1 \,]\*\, \S(1)
          \Big)
\nn\\[-1mm]&&\hspn\vphantom{\FR{1}{1}}
       + \colourcolour{\cfs \*\, \camcf}  \*\,  \Big(
          15 \*\, D_{-1}
          - 9 \*\, D_0
          + \FR{9}{2} \*\, D_1
          + \Big[ \FR{33}{2} \*\, D_1 - 24 \*\, D_{-1} - 6 \*\, D_0 
            + 18 \*\, D_0^2 + 9 \*\, D_1^2 \,\Big] \* \S(1)
\nn\\[-1mm]&&\hspn\vphantom{\FR{1}{1}}
          + [\, 12 \*\, D_0 - 12 \*\, D_{-1} - 6 \*\, D_1 \,]\*\, \S(2)
          + [\, 12 \*\, D_{-1} - 12 \*\, D_0 + 6 \*\, D_1 \,] \*\, \S(-2)
          + [\, 24 \*\, D_{-1} - 24 \*\, D_0
\nn\\[-1mm]&&\hspn\vphantom{\FR{1}{1}}
          + 12 \*\, D_1 \,] \*\, \Ss(1,1)
          \Big)
       + \colourcolour{\cft}  \*\,  \Big(
          10 \*\, D_{-1}
          + 12 \*\, D_{-1}^2
          + 101 \*\, D_0
          - 50 \*\, D_0^2
          + 12 \*\, D_0^3
          - \FR{185}{2} \*\, D_1
          - 34 \*\, D_1^2
\nn\\[-1mm]&&\hspn\vphantom{\FR{1}{1}}
          - 6 \*\, D_1^3
          + 4 \*\, D_2
          + [\, 26 \*\, D_0 - 13 \*\, D_{-1} - 12 \*\, D_{-1}^2 
            - 18 \*\, D_0^2 - \FR{53}{2} \*\, D_1 - 9 \*\, D_1^2 
            - 4 \*\, D_2 \,] \*\, \S(1)
\nn\\[-1mm]&&\hspn\vphantom{\FR{1}{1}}
          + [\, 12 \*\, D_0
          - 12 \*\, D_{-1} - 6 \*\, D_1 \,] \*\, \S(-2)
          \Big)
\Big]
\:\:, \\[2mm]
  \label{ggg3z4}
  \wg_{\,\rm gg}^{\:(3)} &\!\!=\!\!&\vphantom{\FR{1}{1}}
16\*\,\colourcolour{\camcf} \*\, \Big[
       \colourcolour{\cfs\*\,\nf} \*\, \Big(
          95\*\,\eta
          + 54\*\,\eta^2
          + 18\*\,\eta^3
          - 138\*\,\nu
          + 72\*\,\nu^2
          + 11\*\,\S(1)
          \Big)
\nn\\[-1mm]&&\hspn\vphantom{\FR{1}{1}}
       +\, \colourcolour{\camcf\*\,\nfs} \*\, \Big(
          12\*\,\nu
          - 8\*\,\eta
          - 6\*\,\eta^2
          - 2\*\,\S(1)
          \Big)
       + \colourcolour{\camcf^2\*\,\nf} \*\, \Big(
          44\*\,\eta
          + 33\*\,\eta^2
          - 66\*\,\nu
\nn\\[-1mm]&&\hspn\vphantom{\FR{1}{1}}
          + 11\*\,\S(1)
          \Big)
       +\, \colourcolour{\cf\*\,\nfs} \*\, \Big(
          20\*\,\nu
          - 11\*\,\eta
          - 4\*\,\eta^2
          - 2\*\,\S(1)
          \Big)
       +\, \colourcolour{\cf\*\,\camcf\*\,\nf} \*\, \Big(
          \FR{239}{2}\*\,\eta
          + 69\*\,\eta^2
\nn\\[-1mm]&&\hspn\vphantom{\FR{1}{1}}
          + 12\*\,\eta^3
          - 180\*\,\nu
          + 72\*\,\nu^2
          + [\, 22
          - 24\*\,\nu
          + 18\*\,\eta
          + 12\*\,\eta^2 \,]\*\,\S(1)
          \Big)
\Big]
\:\: .
\eea
The abbreviations $\eta$ and $D_k$ have been defined below eq.~(\ref{gnspm3z4});
in addition $\nu \equiv D_{-1} - D_{2}$ is used in eqs.~(\ref{gps3z4}) and
(\ref{ggg3z4}).
$\wg_{\,\rm qq}^{\:(3)}$ is obtained by adding $\,\wg_{\,\rm ns}^{\:+(3)}$
in eq.~(\ref{gnspm3z4}) to the pure-singlet contribution (\ref{gps3z4})
which we were able to check at $N=6$. At least at this value of $N$, we 
expect further checks in the near future. A~complete determination of the
all-$N$ expressions in eqs.~(\ref{gps3z4}) -- (\ref{ggg3z4}) from diagram
calculations, on the other hand, would be a very (currently: too) challenging 
task.

It is interesting to note that all $\z4$ terms of $\gamma^{\:(3)}$ vanish 
for $\cf = \ca$, which is part of the colour-factor choice leading to an 
$\:\!{\cal N}=1$ supersymmetric theory \cite{AntFlo81}. 
The same behaviour has been found (to all orders) before for the 
double-logarithmic large-$N$ contributions to the off-diagonal anomalous 
dimensions $\gamma_{\,\rm qg}$ and $\gamma_{\,\rm gq}$ \cite{ASVxto1}.
We also note that the `diagonal' quantities (\ref{gnspm3z4}), 
(\ref{gps3z4}) and (\ref{ggg3z4}) are reciprocity respecting, see 
ref.~\cite{MRUVV17} and references therein.
 
All flavour-singlet quantities in eqs.~(\ref{gps3z4}) -- (\ref{ggg3z4}) 
include terms with $1/(N-1)^2$ which correspond to $\,\z4\; x^{-1}\ln x\,$ 
terms in the small-$x$ expansion of the N$^3$LO splitting functions 
$P_{\,\rm ik}^{\,(3)}(x)$. 
Contributions of this $x$-space form are also generated, however, by terms 
without $\z4$ in $N$-space. Therefore the above results are not sufficient
to obtain the $\z4$ coefficients of the next-to-next-to-leading 
logarithms of $P_{\,\rm ik}^{\,(3)}(x)$ in the small-$x$ (high-energy) limit.

We now turn to the $\arp5$ (N$^4$LO) contributions to the physical anomalous 
dimensions. Their $\z6$ terms, which we denote by $\widehat{K}^{\:(4)}$, 
are given by simple modifications of eqs.~(\ref{Kns3z4}) and (\ref{K2p3z4}): 
on the right-hand-sides replace $\tilde{f}_{\,\cdots}^{\:(3)}$ everywhere by 
$\hat{f}_{\,\cdots}^{\:(4)}$ for $f = \gamma,\,c$, and replace $-3\:\!\bb0$ 
everywhere by $-4\:\!\bb0$. 
For the non-singlet cases, the $\z4$ terms at N$^4$LO are given by
\beq
\label{Kns4z4}
  \wK_a^{\:(4)} \:=\: \mbox{}
  - \wg_{\,\rm ns}^{\;\sigma(4)} - 3\:\! \bb1\, \wc_a^{\:(3)} - 4\:\! \bb0 
    \left( \wc_a^{\:(4)} - c_a^{(1)}\, \wc_a^{\:(3)} \right)
\eeq
with $\sigma = +$ for $\,a = 2,\rm ns\,$ and $\sigma = -$ for $\,a = 3\,$.
At this order, the scheme transformation of ref.~\cite{Cscheme} includes 
the $\z4$ term of the five-loop beta function, see eq.~(4) of 
ref.~\cite{JM-no-pi2} (where this coefficient, $\bb4$ in our notation, is
denoted by $\bb5$). Consequently, the conjecture of ref.~\cite{JM-no-pi2} 
implies
\bea
\label{Kns4z4a}
  \bb0\,\overline{K}_a^{\:(4)} \:=\: \bb0\, \wK_a^{\:(4)} 
  + \FR13\: \widetilde{\beta}_{\,4} \, K_a^{\,(0)} \;=\; 0 \; .
\eea
Checking this prediction and its $\z6$ counterpart $\,\widehat{K}_a^{\:(4)}=0\,$
requires the four-loop coefficient functions for $F_{\,2,\rm ns}$ and 
$F_{\,3}$ and the corresponding five-loop splitting functions. 
The former quantities have been computed so far at $N \leq 6$ 
\cite{avLL2016,RUVVprp}. The latter have been obtained, very recently, at 
$N=2$ and $N=3$ \cite{RadcTalk,HMRUVVprp} by using the R$^{\ast}$-operation 
\cite{Rstar} as extended to generic numerators in ref.~\cite{RstHR17} and 
implemented using the latest version \cite{Form4.2} of {\sc Form} \cite{FORM}, 
together with the {\sc Forcer} program \cite{Forcer}.
The leading large-$\nf$ contributions to $\gamma_{\,\rm ns}^{\,(4)}$ and 
$c_{\,2,\rm ns}^{\,(4)}$, which both include $\z4$ terms, have been 
determined at all $N$ in refs.~\cite{Lnf}. 

We now show, by explicitly writing down all contributions to 
eqs.~(\ref{Kns4z4}) and (\ref{Kns4z4a}), that the latter relation is 
fulfilled for $K_{\,2,\rm ns}$ at $N=2$. 
The corresponding verification for $K_{\,3}$ at $N=3$ is completely analogous, 
but will be suppressed for brevity. For the same reason we do not show the 
(less critical, since the scheme transformation of ref.~\cite{Cscheme} is not 
required) verification for the $\z6$ parts of $K_{2,\rm ns}$ and $K_{3}$ at 
these values of $N$, nor the all-$N$ verification for the $\z4$ part of 
$K_{2,\rm ns}$ in the large-$\nf$ limit.
The recent result for $\wg_{\,\rm ns}^{\;+(4)}(N\!=\!2)$ 
\cite{RadcTalk,HMRUVVprp} is given by
\bea
\label{gns4z4N2}
  \wg_{\,\rm ns}^{\;+(4)} &\!\! = \!\!&
       \FR{1792}{9} \*\, \nfs \*\, \dfourRRnr
       - \FR{512}{9} \*\, \nf \*\, \dfourRAnr
       + \FR{704}{3} \*\, \cf \*\, \dfourAAna
       + \FR{128}{81} \*\, \cf \*\, \nff
\nn\\&&\hspn\vphantom{\FR{1}{1}}
       - \FR{128}{3} \*\, \cfs \*\, \nft
       + \FR{1072}{81} \*\, \cft \*\, \nfs
       + \FR{21248}{81} \*\, \cff \*\, \nf
       - \FR{10912}{9} \*\, \ca \*\, \dfourRAnr
\nn\\&&\hspn\vphantom{\FR{1}{1}}
       - \FR{5632}{9} \*\, \ca \*\, \nf \*\, \dfourRRnr
       + \FR{2752}{81} \*\, \ca \*\, \cf \*\, \nft
       + \FR{20752}{27} \*\, \ca \*\, \cfs \*\, \nfs
       + \FR{48256}{81} \*\, \ca \*\, \cft \*\, \nf
\nn\\&&\hspn\vphantom{\FR{1}{1}}
       - \FR{59840}{81} \*\, \ca \*\, \cff
       - 784 \*\, \cas \*\, \cf \*\, \nfs
       - \FR{114536}{27} \*\, \cas \*\, \cfs \*\, \nf
       - \FR{229472}{81} \*\, \cas \*\, \cft
\nn\\[1mm]&&\hspn\vphantom{\FR{1}{1}}
       + \FR{274768}{81} \*\, \cath \*\, \cf \*\, \nf
       + \FR{170968}{27} \*\, \cath \*\, \cfs
       - \FR{221920}{81} \*\, \cafo \*\, \cf
\; .
\eea
Here $T_f = 1/2$ has been inserted; the power of $T_f$ for each term can be 
readily reconstructed. 
The result for QCD is obtained for $\,\ca = n_R^{} = 3$, $\,n_A^{} = 8$, 
$\,\cf = 4/3$, $\,d_A^{\,abcd}d_A^{\,abcd}/n_A^{} = 135/8$, 
$\,d_F^{\,abcd}d_A^{\,abcd}/n_{R}^{} = 5/2$ and
$\,d_F^{\,abcd}d_F^{\,abcd}/n_R^{} = 5/36$.
The $\z4$ parts of the four-loop coefficient function 
$c_{\,2,\rm ns}^{\,(4)}(N\!=\!2)$ \cite{avLL2016,RUVVprp} and of the 
five-loop beta function $\bb4$ \cite{beta4} read 
\bea
  \tilde{c}_{\,2,\rm ns}^{\,(4)} &\!\! = \!\!&
       \FR{248}{3} \*\, \dfourRAnr
       + \FR{128}{3} \*\, \nf \*\, \dfourRRnr
       + \FR{16}{27} \*\, \cf \*\, \nft
       - 16 \*\, \cfs \*\, \nfs
       - \FR{220}{27} \*\, \cft \*\, \nf
\nn\\&&\hspn\vphantom{\FR{1}{1}}
       + \FR{1552}{27} \*\, \cff
       + 16 \*\, \ca \*\, \cf \*\, \nfs
       + 176 \*\, \ca \*\, \cfs \*\, \nf
       + \FR{3592}{27} \*\, \ca \*\, \cft
       - \FR{505}{3} \*\, \cas \*\, \cf \*\, \nf
\nn\\[1mm]&&\hspn\vphantom{\FR{1}{1}}
       - 354 \*\, \cas \*\, \cfs
       + \FR{4367}{27} \*\, \cath \*\, \cf
\eea
and
\bea
  \widetilde{\beta}_{\,4} &\!\! = \!\!&
       176 \*\, \nf \*\, \dfourAAna
       - 416 \*\, \nfs \*\, \dfourRAna
       + 128 \*\, \nft \*\, \dfourRRna
       - \FR{44}{3} \*\, \cfs \*\, \nft
\nn\\&&\hspn\vphantom{\FR{1}{1}}
       - 968 \*\, \ca \*\, \dfourAAna
       + 2288 \*\, \ca \*\, \nf \*\, \dfourRAna
       - 704 \*\, \ca \*\, \nfs \*\, \dfourRRna
\nn\\&&\hspn\vphantom{\FR{1}{1}}
       + \FR{28}{3} \*\, \ca \*\, \cf \*\, \nft
       + \FR{286}{3} \*\, \ca \*\, \cfs \*\, \nfs
       + \FR{14}{3} \*\, \cas \*\, \nft
       - \FR{236}{3} \*\, \cas \*\, \cf \*\, \nfs
       - \FR{242}{3} \*\, \cas \*\, \cfs \*\, \nf
\nn\\[1mm]&&\hspn\vphantom{\FR{1}{1}}
       - \FR{26}{3} \*\, \cath \*\, \nfs
       + \FR{451}{3} \*\, \cath \*\, \cf \*\, \nf
       - \FR{583}{6} \*\, \cafo \*\, \nf
       + \FR{121}{6} \*\, \cafi
\; .
\eea
Due to eqs.~(\ref{Kns3z4}) and (\ref{Knsz4=0}), the three-loop contribution
$\tilde{c}_{\,2,\rm ns}^{\,(3)}(N\!=\!2)$ \cite{3loopN2} can be read off from 
Eq.~(\ref{gnspm3z4}). 
For~the convenience of the reader, we also recall the required one-loop 
quantities in the normalization used in this letter:
$\,K_{2,\,\rm ns}^{\,(0)}(N\!=\!2) = \gamma_{\,\rm ns}^{\,(0)}(N\!=\!2) = 
8/3\:\cf\,$,
$\,c_{\,2,\rm ns}^{\,(1)}(N\!=\!2) = 1/3\:\cf\,$ and
$\,\bb0 = 11/3\: \ca - 2/3\:\nf\,$. 
Assembling these contributions, we arrive at eq.~(\ref{Kns4z4a}). This and
the other verifications mentioned above provide substantial and highly 
non-trivial extra evidence for the no-$\pip2$ conjecture in the form presented 
in ref.~\cite{JM-no-pi2}. 
The $\z6$ coefficient of $\gamma_{\:\rm ns}^{\;+(4)}$ at $N=2$ reads
\bea
\label{gns4z6N2}
  \widehat{\gamma}_{\:\rm ns}^{\;+(4)} &\!\! = \!\!&
     \FR{800}{27}\,\*\bb0 \*\, \Big(
         36\, \* \cff
       - 36\, \* \ca\, \* \cft
       - 42\, \* \cas\, \* \cfs
       + 38\, \* \cath\, \* \cf
       - 48\, \* \dfourRAnr
\nn\\[-1mm]&&\vphantom{\FR{1}{1}}
       + 18\, \* \nf\, \* \cft
       - 3\, \* \nf\, \* \ca\, \* \cfs
       - 14\, \* \nf\, \* \cas\, \* \cf
       + 24\, \* \nf\, \* \dfourRRnr
     \,\Big)
\; .
\eea

The $N=4$ expressions corresponding to eqs.~(\ref{gns4z4N2}) and 
(\ref{gns4z6N2}) represent the first new results for the N$^4$LO non-singlet
anomalous dimensions obtained from this conjecture.  They are given by
\bea
  \wg_{\:\rm ns}^{\;+(4)} &\!\! = \!\!&
       - \FR{ 79776202 }{ 50625 }\, \* \ca\, \* \cff
       + \FR{ 6688679 }{ 162000 }\, \* \cas\, \* \cft
       + \FR{ 60495779 }{ 33750 }\, \* \cath\, \* \cfs
       - \FR{ 202467481 }{ 162000 }\, \* \cafo\, \* \cf
\nn\\[1mm]&&\hspn\vphantom{\FR{1}{1}}
       + \FR{ 6908 }{ 15 }\, \* \cf\, \* \dfourAAna
       - \FR{ 349624 }{ 45 }\, \* \ca\, \* \dfourRAnr
       + \FR{ 20332714 }{ 50625 }\, \* \nf\, \* \cff
       - \FR{ 41674913 }{ 81000 }\, \* \nf\, \* \ca\, \* \cft
\nn\\&&\hspn\vphantom{\FR{1}{1}}
       - \FR{ 330775451 }{ 67500 }\, \* \nf\, \* \cas\, \* \cfs
       + \FR{ 428850767 }{ 81000 }\, \* \nf\, \* \cath\, \* \cf
       - \FR{ 124036 }{ 75 }\, \* \nf\, \* \ca\, \* \dfourRRnr
\nn\\&&\hspn\vphantom{\FR{1}{1}}
       + \FR{ 39076 }{ 45 }\, \* \nf\, \* \dfourRAnr
       + \FR{ 1704086 }{ 10125 }\, \* \nfs\, \* \cft
       + \FR{ 8505499 }{ 6750 }\, \* \nfs\, \* \ca\, \* \cfs
       - \FR{ 4882673 }{ 3375 }\, \* \nfs\, \* \cas\, \* \cf 
\nn\\&&\hspn\vphantom{\FR{1}{1}}
       + \FR{ 11704 }{ 25 }\, \* \nfs\, \* \dfourRRnr
       - \FR{ 2146 }{ 25 }\, \* \nft\, \* \cfs
       + \FR{ 139286 }{ 2025 }\, \* \nft\, \* \ca\, \* \cf
       + \FR{ 1256 }{ 405 }\, \* \nff\, \* \cf
\:\: , \\[3mm]
  \widehat{\gamma}_{\:\rm ns}^{\;+(4)} &\!\! = \!\!&
         \FR{ 138248 }{ 27 }\, \* \ca\, \* \cff
       - \FR{ 181522 }{ 27 }\, \* \cas\, \* \cft
       + \FR{ 90475 }{ 54 }\, \* \cath\, \* \cfs
       + \FR{ 102553 }{ 81 }\, \* \cafo\, \* \cf
\nn\\[1mm]&&\hspn\vphantom{\FR{1}{1}}
       + \FR{ 433774 }{ 27 }\, \* \dfourRAnr\, \* \ca
       - \FR{ 25136 }{ 27 }\, \* \nf\, \* \cff
       + \FR{ 136624 }{ 27 }\, \* \nf\, \* \ca\, \* \cft
       - \FR{ 25495 }{ 27 }\, \* \nf\, \* \cas\, \* \cfs
\nn\\&&\hspn\vphantom{\FR{1}{1}}
       - \FR{ 86398 }{ 27 }\, \* \nf\, \* \cath\, \* \cf
       + \FR{ 148016 }{ 27 }\, \* \nf\, \* \dfourRRnr\, \* \ca
       - \FR{ 78868 }{ 27 }\, \* \nf\, \* \dfourRAnr
       - \FR{ 6280 }{ 9 }\, \* \nfs\, \* \cft
\nn\\&&\hspn\vphantom{\FR{1}{1}}
       + \FR{ 3140 }{ 27 }\, \* \nfs\, \* \ca\, \* \cfs
       + \FR{ 43736 }{ 81 }\, \* \nfs\, \* \cas\, \* \cf
       - \FR{ 26912 }{ 27 }\, \* \nfs\, \* \dfourRRnr
\;\; .
\eea
It is also possible to predict the $\z4$ and $\z6$ coefficients 
$\wg_{\,\rm ij}^{\:(4)}$ and $\widehat{\gamma}_{\,\rm ij}^{\:(4)}$
of the N$^4$LO singlet anomalous dimensions at $N=2$ and $N=4$. 
The prediction at $N=2$ includes a further check, since the four results must 
be pairwise equal due to the momentum sum rule. By evaluating the elements
of $K$ in eq.~(\ref{F2Fphi}) as given by eq.~(\ref{Kdef}), we obtain
\bea
  \label{gqq4N2z4}
    \wg_{\,\rm qq}^{\:(4)}(N\!=\!2) &\!\!=\!\!& \mbox{}
  - \wg_{\,\rm gq}^{\:(4)}(N\!=\!2)
\nn \\[1mm]
  &\!\!=\!\!&
       \FR{2048}{9} \*\, \nfs \*\, \dfourRRnr
       - \FR{512}{9} \*\, \nf \*\, \dfourRAnr
       + \FR{704}{3} \*\, \cf \*\, \dfourAAna
       + \FR{640}{81} \*\, \cf \*\, \nff
\nn\\&&\hspn\vphantom{\FR{1}{1}}
       - \FR{10496}{81} \*\, \cfs \*\, \nft
       + \FR{2896}{27} \*\, \cft \*\, \nfs
       + \FR{1024}{3} \*\, \cff \*\, \nf
       - \FR{10912}{9} \*\, \ca \*\, \dfourRAnr
\nn\\&&\hspn\vphantom{\FR{1}{1}}
       - \FR{7040}{9} \*\, \ca \*\, \nf \*\, \dfourRRnr
       + \FR{6188}{81} \*\, \ca \*\, \cf \*\, \nft
       + \FR{142996}{81} \*\, \ca \*\, \cfs \*\, \nfs
\nn\\&&\hspn\vphantom{\FR{1}{1}}
       + \FR{6596}{9} \*\, \ca \*\, \cft \*\, \nf
       - \FR{59840}{81} \*\, \ca \*\, \cff
       - \FR{48598}{27} \*\, \cas \*\, \cf \*\, \nfs
       - \FR{196028}{27} \*\, \cas \*\, \cfs \*\, \nf
\nn\\[1mm]&&\hspn\vphantom{\FR{1}{1}}
       - \FR{229472}{81} \*\, \cas \*\, \cft
       + 6254 \*\, \cath \*\, \cf \*\, \nf
       + \FR{170968}{27} \*\, \cath \*\, \cfs
       - \FR{221920}{81} \*\, \cafo \*\, \cf
\:\: ,
\\[2mm]
  \label{gqg4N2z4}
    \wg_{\,\rm qg}^{\:(4)}(N\!=\!2) &\!\!=\!\!& \mbox{}
  - \wg_{\,\rm gg}^{\:(4)}(N\!=\!2)
\nn \\[1mm]
  &\!\!=\!\!&
       \FR{256}{9} \*\, \nfs \*\, \dfourRAna
       - \FR{176}{3} \*\, \nf \*\, \dfourAAna
       - \FR{1024}{9} \*\, \nft \*\, \dfourRRna
       + \FR{512}{81} \*\, \cf \*\, \nff
\nn\\&&\hspn\vphantom{\FR{1}{1}}
       - \FR{3844}{81} \*\, \cfs \*\, \nft
       + \FR{3808}{81} \*\, \cft \*\, \nfs
       + \FR{6400}{81} \*\, \cff \*\, \nf
       + \FR{5456}{9} \*\, \ca \*\, \nf \*\, \dfourRAna
\nn\\&&\hspn\vphantom{\FR{1}{1}}
       + \FR{3520}{9} \*\, \ca \*\, \nfs \*\, \dfourRRna
       - \FR{176}{81} \*\, \ca \*\, \nff
       + \FR{4928}{81} \*\, \ca \*\, \cf \*\, \nft
       + \FR{40348}{81} \*\, \ca \*\, \cfs \*\, \nfs
\nn\\&&\hspn\vphantom{\FR{1}{1}}
       + \FR{5932}{27} \*\, \ca \*\, \cft \*\, \nf
       - \FR{910}{27} \*\, \cas \*\, \nft
       - \FR{11360}{9} \*\, \cas \*\, \cf \*\, \nfs
       - \FR{136090}{81} \*\, \cas \*\, \cfs \*\, \nf
\nn\\[1mm]&&\hspn\vphantom{\FR{1}{1}}
       + \FR{60019}{81} \*\, \cath \*\, \nfs
       + \FR{313570}{81} \*\, \cath \*\, \cf \*\, \nf
       - \FR{204065}{81} \*\, \cafo \*\, \nf
\eea
and
\bea
  \label{gqq4N2z6}
    \widehat{\gamma}_{\,\rm qq}^{\:(4)}(N\!=\!2) &\!\!=\!\!& \mbox{}
  - \widehat{\gamma}_{\,\rm gq}^{\:(4)}(N\!=\!2) 
\nn \\[1mm]
  &\!\!=\!\!&
       \FR{25600}{27} \*\, \nf \*\, \dfourRAnr
       - \FR{25600}{27} \*\, \nfs \*\, \dfourRRnr
       - \FR{8000}{9} \*\, \cft \*\, \nfs
       - \FR{3200}{3} \*\, \cff \*\, \nf
\nn\\&&\hspn\vphantom{\FR{1}{1}}
       - \FR{140800}{27} \*\, \ca \*\, \dfourRAnr
       + \FR{140800}{27} \*\, \ca \*\, \nf \*\, \dfourRRnr
       + \FR{4000}{27} \*\, \ca \*\, \cfs \*\, \nfs
\nn\\&&\hspn\vphantom{\FR{1}{1}}
       + \FR{139600}{27} \*\, \ca \*\, \cft \*\, \nf
       + \FR{35200}{9} \*\, \ca \*\, \cff
       + \FR{56800}{81} \*\, \cas \*\, \cf \*\, \nfs
       + \FR{10600}{27} \*\, \cas \*\, \cfs \*\, \nf
\nn\\[1mm]&&\hspn\vphantom{\FR{1}{1}}
       - \FR{35200}{9} \*\, \cas \*\, \cft
       - 4200 \*\, \cath \*\, \cf \*\, \nf
       - \FR{123200}{27} \*\, \cath \*\, \cfs
       + \FR{334400}{81} \*\, \cafo \*\, \cf
\:\: ,
\\[2mm]
  \label{ggq4N2z6}
    \widehat{\gamma}_{\,\rm qg}^{\:(4)}(N\!=\!2) &\!\!=\!\!& \mbox{}
  - \widehat{\gamma}_{\,\rm gg}^{\:(4)}(N\!=\!2)
\nn \\[1mm]
  &\!\!=\!\!&
       \FR{12800}{27} \*\, \nft \*\, \dfourRRna
       - \FR{12800}{27} \*\, \nfs \*\, \dfourRAna
       - \FR{3200}{9} \*\, \cft \*\, \nfs
       - \FR{3200}{9} \*\, \cff \*\, \nf
\nn\\&&\hspn\vphantom{\FR{1}{1}}
       + \FR{70400}{27} \*\, \ca \*\, \nf \*\, \dfourRAna
       - \FR{70400}{27} \*\, \ca \*\, \nfs \*\, \dfourRRna
       + 400 \*\, \ca \*\, \cfs \*\, \nfs
\nn\\&&\hspn\vphantom{\FR{1}{1}}
       + \FR{41200}{27} \*\, \ca \*\, \cft \*\, \nf
       + \FR{800}{81} \*\, \cas \*\, \nft
       + \FR{1400}{9} \*\, \cas \*\, \cf \*\, \nfs
       - \FR{16400}{9} \*\, \cas \*\, \cfs \*\, \nf
\nn\\[1mm]&&\hspn\vphantom{\FR{1}{1}}
       - \FR{7400}{27} \*\, \cath \*\, \nfs
       - \FR{12100}{27} \*\, \cath \*\, \cf \*\, \nf
       + \FR{97900}{81} \*\, \cafo \*\, \nf
\:\: .
\eea
Hence the no-$\pip2$ conjecture also passes this further five-loop check. 
We note that this check succeeds only due to the ($\z4$ part of the) scheme 
transformation of ref.~\cite{Cscheme}. 
This is not due to the $(F_{\,2},\,F_{\,\phi})$ analogue of the 
$\widetilde{\beta}_4$ shift (\ref{Kns4z4a}), as the resulting 
contributions to the momentum sum rule cancel. 
Instead it arises from the need to refer to a renormalization-group invariant
current \cite{JM-no-pi2}, which is not $G^{\,\mu\nu} G_{\!\mu\nu}$ but 
$\,\beta(\ar)/\ar \; G^{\,\mu\nu} G_{\!\mu\nu}$ for the structure function 
$F_\phi$. 
The resulting overall factor of $\ar$ induces a scheme shift $\,\sim 
\widetilde{\beta}_4/\bb0\; c_{\phi,\rm g}^{\:(0)}$ of the N$^4$LO coefficient 
function $\wc_{\phi,\rm g}^{\:(4)}$, with $c_{\phi,\rm g}^{\:(0)} = 1$.

Finally we step back to N$^3$LO and address the longitudinal structure
function $F_{\,L}$. The physical anomalous dimensions for $F_{\,L,\rm ns}$
and the singlet system $(F_{\,2}, F_{\,L})$ \cite{CataniF2FL}, see also 
ref.~\cite{BRvNphys}, have been employed in refs.~\cite{physFL1} to predict
large-$N$ double logarithms.
It is convenient to consider ${\cal F}_L = F_{\:\!L} / (\ar\, c_{L,q}^{(1)})$ 
with the coefficient functions 
 $c_{\lambda,\rm i}^{\,(3)} \equiv c_{L,\rm i}^{\,(4)} / c_{L,q}^{(1)}$
with $c_{L,q}^{(1)} = 4\,\cf / (N\!+\!1)$ -- recall our normalization 
$\ar=\as/4 \pi\,$ of the reduced coupling. 

The non-singlet case is then directly analogous to eqs.~(\ref{Kns3z4}), hence 
$\wK_{\,L,\rm ns} = 0$ together with eq.~(\ref{gnspm3z4}) leads to an all-$N$ 
prediction that we have checked against diagram calculations at $N=2$, $N=4$ 
and $N=6$ \cite{avLL2016,RUVVprp}. This prediction reads
\bea
  \label{cLns4z4}
  \!\!\!\hspn\wc_{\,L,\,\rm ns}^{\:(4)} &\!\!\!\!=\!\!\!\!&\vphantom{\FR{1}{1}}
16\*\,\colourcolour{\cfs \*\, \camcf} \*\, D_1 \* \Big[
        6 \*\, \colourcolour{\camtcf} \*\, \Big(
          ( \eta + \eta^2 )
          - \FR{5}{4}
          - 2 \*\, \S(-2)
          \Big)
       + \colourcolour{\nf} \*\, \Big(
          3
          + 2 \*\, \eta
          - 4 \*\, \S(1)
          \Big)
\Big]
\:\: \!\!.
\eea

The structure of the corresponding anomalous-dimension matrix for 
$(F_{\,2}, {\cal F}_{\,L})$ is more involved than that in eqs.~(\ref{K2p3z4}) 
for $(F_{\,2}, F_{\,\phi})$, since the leading-order analogue $C_\lambda$ of 
$C$ in eq.~(\ref{F2Fphi}) is not given by the unit matrix, but by
\beq
\label{C2lam}
  C_\lambda^{\,(0)} \:=\:
  \left(  \begin{array}{cc} 1 & 0 \\
         1 & C_{\lambda,\,g}^{\,(0)} \end{array} \right)
\:\: .
\eeq
Nevertheless, it is of course no problem to evaluate eq.~(\ref{Kdef}) by
symbolic manipulation to N$^3$LO accuracy also in this case. 
We have checked that the $(F_{\,2}, {\cal F}_{\,L})$ analogues of 
eq.~(\ref{K2p3z4}), here suppressed for brevity, lead to 
\beq
\label{KLijz4=0}
  \wK_{22}^{\:(3)} \:=\: \wK_{2L}^{\:(3)} \:=\:
  \wK_{L2}^{\:(3)} \:=\: \wK_{LL}^{\:(3)} \:=\: 0
\eeq
for $N=2$ and $N=4$. Note that $K_{22}$ here is not the same as $K_{22}$
in eq.~(\ref{K2p3z4}). 

The all-$N$ forms of four-loop quantities $\wc_{L,\rm ps}^{\:(4)}$ 
and $\wc_{L,\rm g}^{\:(4)}$ can be predicted by imposing eq.~(\ref{KLijz4=0})
at all even-$N$ and using eqs.~(\ref{gps3z4})$\,$--$\,$(\ref{ggg3z4}) for the 
four-loop splitting functions. In this manner we arrive~at
\bea
  \label{cLps4z4}
  \wc_{\,L,\,\rm ps}^{\:(4)} &\!\!=\!\!&\vphantom{\FR{1}{1}}
16\*\,\colourcolour{\cf \*\, \camcf} \*\, \Big[
       \colourcolour{\nfs}  \*\,  \Big(
          \FR{8}{3} \*\, D_{-1}
          - 8 \*\, D_0
          + 8 \*\, D_1^2
          + \FR{16}{3} \*\, D_2
          \Big)
       + \colourcolour{\camtcf \*\, \nf}  \*\,  \Big(
          4 \*\, D_{-1}
\nn\\[-1mm]&&\hspn\vphantom{\FR{1}{1}}
          + 30 \*\, D_0
          - 12 \*\, D_0^2
          - 42 \*\, D_1
          - 18 \*\, D_1^2
          - 12 \*\, D_1^3
          + 8 \*\, D_2
          - 4 \*\, [ D_{-1} - 3 \*\, D_0 + 3 \*\, D_1^2 + 2 \*\, D_2] \*\, \S(1)
          \Big)
\nn\\[-1mm]&&\hspn\vphantom{\FR{1}{1}}
       + \colourcolour{\cf \*\, \nf}  \*\,  \Big(
          30 \*\, D_1
          + 6 \*\, D_1^2
          + 12 \*\, D_1^3
          - 4 \*\, D_{-1}
          - 18 \*\, D_0
          + 12 \*\, D_0^2
          - 8 \*\, D_2
          \Big)
\Big]
\:\:,\\[3mm]
  \label{cLgg4z4}
  \wc_{\,L,\,\rm g\hphantom{s}}^{\:(4)} &\!\!=\!\!&\vphantom{\FR{1}{1}}
16\*\,\colourcolour{\camcf} \*\, \Big[
       \colourcolour{\camcf \*\, \nfs}  \*\, \Big(
          \FR{8}{3} \*\, D_{-1}
          + 26 \*\, D_0
          - 12 \*\, D_0^2
          - 72 \*\, D_1
          + 8 \*\, D_1^2
          - 24 \*\, D_1^3
          + \FR{130}{3} \*\, D_2
\nn\\[-1mm]&&\hspn\vphantom{\FR{1}{1}}
          + 16 \*\, D_2^2
          - 8 \*\, [ D_1 - D_2] \*\, \S(1)
          \Big)
       + \colourcolour{\cf \*\, \nfs}  \*\, \Big(
          \FR{8}{3} \*\, D_{-1}
          - 10 \*\, D_0
          - 24 \*\, D_1
          + 20 \*\, D_1^2
          + \FR{94}{3} \*\, D_2
          + 16 \*\, D_2^2
\nn\\[-1mm]&&\hspn\vphantom{\FR{1}{1}}
          - 8 \*\, [ D_1 - D_2] \*\, \S(1)
          \Big)
       + \colourcolour{\cf \*\, \camcf \*\, \nf}  \*\, \Big(
          117 \*\, D_1
          - 12 \*\, D_1^2
          + 36 \*\, D_1^3
          - 4 \*\, D_{-1}
          - 39 \*\, D_0
          + 18 \*\, D_0^2
\nn\\[-1mm]&&\hspn\vphantom{\FR{1}{1}}
          - 74 \*\, D_2
          - 24 \*\, D_2^2
          + [6 \*\, D_0 + 6 \*\, D_1 - 12 \*\, D_1^2 - 12 \*\, D_2] \*\, \S(1)
          \Big)
       + \colourcolour{\cfs \*\, \nf}  \*\, \Big(
          84 \*\, D_1
          - 48 \*\, D_1^2
          + 24 \*\, D_1^3
\nn\\[-1mm]&&\hspn\vphantom{\FR{1}{1}}
          - 4 \*\, D_{-1}
          - 6 \*\, D_0
          + 12 \*\, D_0^2
          - 74 \*\, D_2
          - 24 \*\, D_2^2
          + [6 \*\, D_0 + 6 \*\, D_1 - 12 \*\, D_1^2 - 12 \*\, D_2] \*\, \S(1)
          \Big)
\Big]
\:\: .
\eea
The flavour-singlet quark coefficient function $\wc_{L,\rm q}^{\:(4)}$ is 
obtained by adding $\wc_{L,\rm ns}^{\:(4)}$ in eq.~(\ref{cLns4z4}) to the
pure-singlet quantity (\ref{cLps4z4}). Since only these two coefficient 
functions have been determined, two of the four relations in 
eq.~(\ref{KLijz4=0}) are left as additional all-$N$ checks of the no-$\pip2$ 
conjecture.

To summarize: we have presented a large amount of additional evidence for 
the conjecture that there are no $\pip2$ contributions to the expansion 
coefficients of Euclidean physical quantities in massless perturbative QCD 
and its generalization to a general simple compact group group
\cite{JM-no-pi2}. 
Besides low even or odd integer-$N$ values of the physical anomalous 
dimensions for the non-singlet structure functions $F_{a,\rm ns}$, 
$a = 2,3,L$, and the singlet systems $(F_{\,2},\,F_{\,\phi})$,
$(F_{\,2},\,F_{\,L})$ at N$^3$LO and N$^4$LO \linebreak
-- in~the latter case the conjecture holds only after the scheme transformation 
 of ref.~\cite{Cscheme}, or indeed after any scheme transformation that 
 removes the $\z4$ contributions to $\bb4$ occurring in \MSb\ \cite{beta4} --
these checks include four `all-$N$' relations at N$^3$LO, three for even $N$ 
and one for odd $N$.

Based on the evidence presented in ref.~\cite{JM-no-pi2} and in this letter,
this conjecture can be employed to predict new $\pip2$ contributions to 
higher-order anomalous dimensions and coefficient functions.
At N$^3$LO we have presented the $\z4$ contributions to the flavour-singlet
splitting functions and to the coefficient functions for the longitudinal
structure function $F_{\,L}$ at all even $N$.
Based on present four-loop {\sc Forcer} \cite{Forcer} computations of 
DIS~\cite{avLL2016,RUVVprp}, it is possible to predict hitherto unknown 
$\z4$ and $\z6$ parts of N$^4$LO anomalous dimensions at $N \leq 6$. Here
we have shown, for brevity, only the $N=2$ ($N=4$) results for the 
singlet (non-singlet) case. 
These predictions, and the no-$\pip2$ conjecture in general, will serve as 
useful partial checks for very complicated future high-order computations. 
They may also provide input for future studies of the structure of 
perturbative quantum field theory.

 
{\small
\setlength{\baselineskip}{0.4cm}

}

\end{document}